# Mobile Ad Hoc Networks:
# A Comparative Study of QoS Routing Protocols

SANJEEV GANGWAR[1], DR. SAURABH PAL[1] and DR. KRISHAN KUMAR[3]

[1,2] Department of Computer Application, VBS Purvanchal University Jaunpur
[1]gangwar.sanjeev@gmail.com
[2]drsaurabhpal@yahoo.co.in
[3] Department of Computer Application, Gurukula Kangri Vishwavidyalaya, Haridwar
[3]kumar.krishana@gmail.com

***ABSTRACT** This Article presents a thorough overview of QoS routing metrics, resources and factors affecting performance of QoS routing protocols. The relative strength, weakness, and applicability of existing QoS routing protocols are also studied and compared. QoS routing protocols are classified according to the QoS metrics used type of QoS guarantee assured.*

## 1. INTRODUCTION

Mobile Ad Hoc Networks (MANETs) is a class of wireless networks that have been researched extensively over the recent years [1]. MANETs do not require the support of wired access points or base stations for intercommunication. A mobile ad hoc network, unlike a static network, has no infrastructure. It is a collection of mobile nodes where communication is established in the absence of any fixed foundation. The only possible direct communication is between neighboring nodes. Therefore, communication between remote nodes is based on multiple-hop. These nodes are dynamically and arbitrarily located in such a manner that the interconnections between nodes are capable of changing on a continual basis. MANETs are self-configuring; there is no central management system with configuration responsibilities. All the mobile nodes can communicate each other directly, if they are in other's wireless links radio range. In order to enable data transfer they either communicate through single hop or through multiple hops with the help of intermediate nodes. Since MANETs allow ubiquitous service access, anywhere, anytime without any fixed infrastructure they can be widely used in military battlefields, crisis management services, classrooms and conference halls etc. MANETs ad-hoc fashion networking developments lead to development of enormous multimedia applications such as video-on-demand, video conferencing etc. Routing in mobile ad hoc networks and some fixed wireless networks use multiple-hop routing. Routing protocols for this kind of wireless network should be able to maintain paths to other nodes and, in most cases, must be handle changes in paths due to mobility. However, most of the existing Ad Hoc routing protocols do not consider the QoS problem. Quality of Service (QoS) is the performance level of a service offered by the network to the user. Most of the multimedia applications have stringent QoS requirements that must be satisfied. The goal of QoS provisioning is to achieve a more deterministic network behavior, so that information carried by the network can be better delivered and network resources can be better utilized. However, there still remains a significant challenge to provide QoS solutions and maintain end-to-end QoS with user mobility. Most of the conventional routing protocols are designed either to minimize the data traffic in the network or to minimize the average hops for delivering a packet. [1]. Even some protocols such as Ad-hoc On demand Distance Vector (AODV) [2], Dynamic Source Routing (DSR) [3] and On-demand Multicast Routing Protocol (ODMRP) [4] are designed without explicitly considering QoS. When QoS is considered, some protocols may be unsatisfactory or impractical due to the lack of resources and the excessive computation overhead. QoS routing usually involves two tasks: collecting and maintaining up-to-date state information about the network and finding feasible paths for a connection based on its QoS requirements. [5] To support QoS, a service can be characterized by a set of measurable pre specified service requirements such as minimum bandwidth, maximum delay, maximum delay variance and maximum packet loss rate. However many other metrics are also used to quantify QoS and in this paper we cover most of them in detail.

The remainder of this paper is organized as follows. In Section 2, we discuss related works in terms of QoS routing surveys and summarize their main points. Review of the several challenges faced by the provision of QoS on the MANET environment is given is section 3. In section 4, we analyze the QoS routing metrics commonly used by all applications and the tradeoffs involved in the protocol design. Section 5 and 6 presents the taxonomy of QoS routing protocols based on their network architecture, type of QoS guarantee assured and the interaction with the MAC layer. Following this, we summarize and compare the operations, key features and major advantages and drawbacks of a selection of QoS routing protocols proposed in the literature. Finally we draw the conclusions and discuss our findings in the field of QoS routing. We ask that authors follow some simple guidelines. In essence, we ask you to make your paper look exactly like this document. The easiest way to do this is simply to download the template, and replace the content with your own material.

## 2. RELATED WORKS

Routing protocols belonging to different QoS philosophies have been proposed in the literature. A fairly comprehensive overview of the state of the field of QoS in networking was provided by Chen in 1999 [6]. Chakrabarti and Mishra [7] later summarized the important QoS related issues in MANETs in 2001 and their conclusions highlighted several significant points in MANET research. It includes admission control policies and protocols, QoS robustness and QoS preservation under failure conditions. In 2004, Al-Karaki and Kamal published a detailed overview [8] and the development trends in the field of QoS routing. They highlighted some areas such as security and multicast routing requiring further research attention. They were categorized the QoS routing solutions into various types of approaches: Flat, Hierarchical, Position-based and power aware QoS routing.Reddy et al. [9] provided a thorough overview of the more widely accepted MAC and routing solutions for providing better QoS in MANETs.

## 3. ISSUES AND CHALLENGES WHILE PROVIDING QOS IN AD-HOC NETWORKS

QoS provision will lead to an increase in computational and communicational cost. In other words, it requires more time to setup a connection and maintains more state information per connection. The improvement in network utilization counterbalances the increase in state information and the associated complexity and various issues are needed to be faced while providing QoS for MANETS. The major problems that are faced are as follows:

**Unreliable channel:** The bit errors are the main problem which arises because of the unreliable wireless channels. These channels cause high bit error rate and this is due to high interference, thermal noise, multipath fading effects, [10] and so on. This leads to low packet delivery ratio. Since the medium is wireless in the case of MANETs, it may also lead to leakage of information into the surroundings.

**Maintenance of route:** The dynamic nature of the network topology and changing behavior of the communication medium makes the maintenance of network state information very difficult. The established routing paths may be broken even during the process of data transfer. Hence the need for maintenance and reconstruction of routing paths with minimal overhead and delay causes. The QoS aware routing would require the reservation of resources at the intermediate nodes. The reservation maintenance with the changes in topology becomes cumbersome.

**Mobility of the node:** Since the nodes considered here are mobile nodes, that is they move independently and randomly at any direction and speed, the topology information has to be updated frequently and accordingly so as to provide routing to reach the final destination which result in again less packet delivery ratio. [11]

**Limited power supply:** The mobile nodes are generally constrained by limited power supply compared to nodes in the wired networks. Providing QoS consumes more power due to overhead from the mobile nodes which may drain the node's power quickly.

**Lack of centralized control:** The members of any ad hoc networks can join or leave the network dynamically and the network is set up spontaneously. So there may not be any provision of centralized control on the nodes which leads to increased algorithm's overhead and complexity, as QoS state information must be disseminated efficiently.

**Channel contention:** Nodes in a MANET must communicate with each other on a common channel so as to provide the network topology. However, this introduces the problems of interference and channel contention. For peer-to-peer data communications these can be avoided in various ways. One way is to attempt global clock synchronization and use a TDMA-based system where each node may transmit at a predefined time. This is difficult to achieve since there is no centralized control on the nodes. Other ways are to use a different frequency band or spreading code (as in CDMA) for each transmitter. This requires a distributed channel selection mechanism as well as the dissemination of channel information[12].

**Security:** Security can be considered as a QoS attribute. Without adequate security, unauthorized accesses and usages may violate the QoS negotiations. The nature of broadcasts in wireless networks potentially results in more security exposures. The physical medium of communication is inherently insecure. So we need to design security-aware routing algorithms for ad hoc networks. Please use a 9-point Times Roman font, or other Roman font with serifs, as close as possible in appearance to Times Roman in which these guidelines have been set. The goal is to have a 9-point text, as you see here. Please use sans-serif or non-proportional fonts only for special purposes, such as distinguishing source code text. If Times Roman is not available, try the font named Computer Modern Roman. On a Macintosh, use the font named Times. Right margins should be justified, not ragged.

## 4. EVALUATION METRICS FOR QOS ROUTING PROTOCOLS

As different applications have different requirements, the services required by them and the associated QoS parameters differ from application to application. For example, in case of multimedia applications, bandwidth, delay and delay-jitter are the key QoS parameters, whereas military applications have stringent security requirements. The following is a sample of the metrics commonly used by applications to specify QoS requirement to the routing protocol.

- Minimum Throughput (bps) – the desired application data throughput. [13]
- Maximum Delay (s) – maximum tolerable end-to-end delay for data packets. [14]
- Maximum Delay jitter – difference between the upper bound on end-to-end delay and the absolute minimum delay. [15]
- Maximum Packet loss ratio - the acceptable percentage of total packets sent, which are not received by the final destination node. [16]

The value of a metric over the entire path can be one of the following compositions [25][26]:

• **Additive metrics:** This can be represented mathematically as follows:

$$m(p) = \sum_{i=1}^{LK} m(lk_i)$$

Where m (p) is the total of metric m of path (p), $lk_i$ is a link in the path (p), LK is the number of links in path (p), and i= 1,…LK Delay, delay variation (jitter), and cost are examples of this type of composition. Various factors that determine the delay in communication networks are reviewed in [23].

• **Concave metrics**: This can be represented mathematically as follows:

$$m(p) = \min(m(lk_i))$$

Bandwidth is an example of this type of composition. The bandwidth we are interested in here is the residual bandwidth that is available for new traffic. It can be defined as the minimum of the residual bandwidth of all links on the path or the bottleneck bandwidth.

- Multiplicative metrics. This can be represented mathematically as follows:

$$m(p) = \prod_{i=1}^{LK} m(lk_i)$$

Loss probability is an indirect example of this type of composition.

- Convex metrics: This can be represented as the maximum of all metric along the path

$$m(p)=\max (m(lk_i))$$

Vulnerability (in context of security) and throughput use the convex rule. Whatever the metrics used in determining the path, these metrics must represent the basic network properties of interest. Such metrics include residual bandwidth, delay, and jitter. Since the flow QoS requirements have to be mapped onto path metrics, the metrics define the types of QoS guarantees the network can support. Alternatively, QoS-based routing cannot support QoS requirements that cannot be meaningfully mapped onto a reasonable combination of path metrics.

## 5. CRITERIA OF QOS ROUTING PROTOCOLS CLASSIFICATION

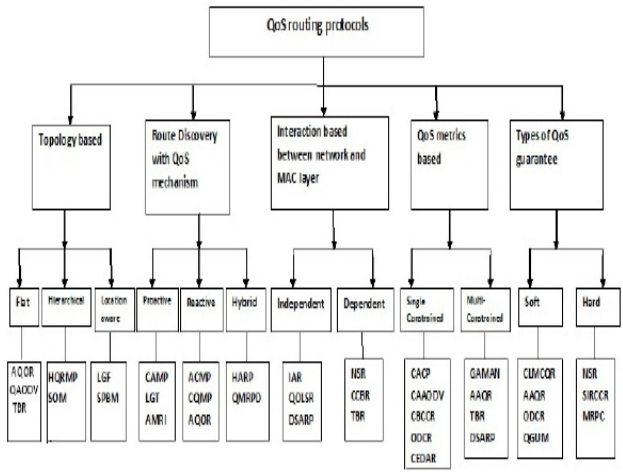

Fig 1: QoS routing protocols classification

**Route Discovery with QoS based protocols**
Based on the routing information update mechanism employed, QoS approaches can be classified into three categories viz., Proactive, on-demand, and hybrid QoS approaches. Proactive protocols are one where a routing table is maintained at every node which aids in forwarding packets. These tables are updated regularly in order to maintain up-to-date routing information from each node to every other node. Therefore, the source node can get a routing path immediately if it needs one. There are some typical proactive QoS routing protocols such as QOLSR [23] (QoS Optimized Link State Routing) and PLBQR [24] (Predictive Location-Based QoS Routing in Mobile Ad Hoc Networks). A reactive protocol is also called "on-demand" protocols. Reactive protocols are one which does not require the maintenance of network topology when there is no traffic. The state information is acquired when needed. However, route maintenance is an important operation of reactive routing protocols, because source nodes may suffer from long delays for route searching before they can forward data packets. QoS AODV [25] (QoS Ad-hoc on demand Distance Vector), ACMP [26] (Adaptive Core based Routing Protocol with Consolidated Query Packets) and CQMP (Mesh-based Multicast Routing Protocol with Consolidated Query Packets) are typical examples for reactive routing protocols. Compared to proactive routing protocols, less control overhead is the significant advantage of the reactive routing protocols. A hybrid protocol as the name implies it is a combination of both proactive and reactive strategies. Hence, hybrid protocols address both efficiency and robustness. The Efficient hybrid Multicast Routing Protocol (EHMRP) [26] is an instance for hybrid-based QoS routing protocol.

**Single constrained vs. Multi constrained QoS metrics**
Most of the protocols focused on providing an assured throughput service only, since Throughput was deemed the most important requirement in earlier days. These single-constrained routing protocols have received success in many aspects; however, they do not always perform best. In CEDAR the bandwidth is used as the only QoS parameter for routing. Most of the multimedia applications require the communication to meet stringent requirements on delay, delay-jitter, cost and other QoS metrics. In this context, the trend is to move from single constrained routing to multi constrained routing. The main function of multiconstrained QoS routing is to find a feasible path that satisfies multiple constraints simultaneously, which is a big challenge for MANETs where the topology may change constantly. It has been proved that such a problem is NP-complete. QMRPD (QoS Multicast Routing Protocol for Dynamic group topology) [33] GAMAN (Genetic Algorithm-based routing for MANETs) [34] HMCOP (Heuristic multi Constrained Optimal Path) are typical multi constrained routing protocols.

**Hard QoS vs. Soft QoS approach**
The QoS provisioning approaches can be broadly classified into two categories, hard QoS and soft QoS approaches. If QoS requirements of a connection are guaranteed to be met for the whole duration of the session, the QoS approach is termed as hard QoS approach. In MANETS it is very challenging to provide hard QoS guarantees to user applications. Some of the protocols NSR and SIRCCR (SIR and Channel Capacity based Routing). If the QoS requirements are not guaranteed for the entire session, the QoS approach is termed as soft QoS approach. Thus, QoS guarantees can only be given within certain statistical bounds. Most of the protocols provide soft QoS guarantees.

## 6. QOS-AWARE ROUTING PROTOCOLS
The primary goal of the QoS-aware routing protocols is to determine a path from a source to the destination that satisfies the needs of the desired QoS. The QoS-aware path is determined within the constraints of bandwidth, minimal search, distance, and traffic conditions. Since path selection is based on the desired QoS, the routing protocol can be termed QoS-aware. In the literature, numerous routing protocols have been proposed for finding QoS paths. In the following sections some of these QoS routing protocols are described.

### 6.1 CEDAR
The Core-Extraction Distributed Ad hoc Routing (CEDAR) algorithm is proposed for QoS routing in ad hoc networks. Bandwidth information is advertised by elected subset nodes along with their link state updates, to identify and avoid congested parts of the network. When a link fails, CEDAR's route re-computation confines itself to the immediate neighborhood of the breakage.

Core extraction: A set of nodes is elected to form the core that maintains the local topology of the nodes in its domain, and also to perform route computations. The core nodes are elected by approximating a minimum dominating set1 of the ad hoc network.

Link state propagation: QoS routing in CEDAR is achieved by propagating the bandwidth availability information of stable links to all core nodes. The basic idea is that the information about stable high bandwidth links can be made known to nodes far away in the network, while information about the dynamic or low bandwidth links remains within the local area.

Route computation: Route computation first establishes a core path from the domain of the source to the domain of the destination. Using the directional information provided by the core path, CEDAR iteratively tries to find a partial route from the source to the domain of the furthest possible node in the core path satisfying the requested bandwidth. This node then becomes the source of the next iteration. In the CEDAR approach, the core provides an efficient low-overhead infrastructure to perform routing, while the state propagation mechanism ensures availability of link state information at the core nodes without incurring high overheads.

### 6.2 Multipath Routing Protocol (MRP)
MRP is a reactive on-demand routing Protocol which extends DSR protocol to find multipath routing coupled with bandwidth and reliability constraint. It consists of three phases: routing discovery, routing maintenance and traffic allocation. In routing discovery phase, the protocol selects several multiple alternate paths which meet the QoS requirements and the ideal number of multipath routing is achieved to compromise between load balancing and network overhead. In routing maintenance phase, it can effectively deal with route failures similar to DSR. Furthermore, the per-packet granularity is adopted in traffic allocation phase.

### 6.3 Genetic Algorithm-Based QoS Routing Protocol for MANETS (GAMAN)
A Genetic Algorithm-based source-routing Protocol for MANETs (GAMAN) is proposed, which uses end-to-end delay and transmission success rate for QoS metrics. Genetic Algorithms (GAs) may be employed for heuristically approximating an optimal solution to a problem, in this case finding the optimal route based on the two QoS constraints mentioned. The first stage of the process involves encoding routes so that a GA can be applied; this is termed gene coding. For this purpose, paths are discovered on-demand and then a network topology view is constructed in a logical tree-like structure. Each node stores a tree routed at itself with its neighbor nodes as child nodes and in turn their neighbor nodes as their children. The route discovery algorithm is assumed to collect locally computed metrics such as average delay over a link and the link reliability for the links on each path. After the gene encoding stage, the fitness T of each path is calculated as follows:

$$T = \frac{\sum_{i=1}^{m} D_i}{\prod_{i=1}^{n} R_i}$$

where $D_i$ and $R_i$ are the delay and reliability of link i, respectably. The fitness values are used to select paths for cross-over breeding and mutation operations. The fittest path (with the smallest T) and the offspring from the genetic operations are carried forward into the next generation. While this method is a useful heuristic for approximating the optimal value over the delay and link reliability metrics at the same time, it requires many paths to be searched in order to collect enough "genetic information" for the GA operations to be meaningful. This means that the method is not suited to large networks

### 6.4 Predictive Location-Based QoS Routing in Mobile Ad Hoc Networks (PLBQR)
It is a location aware QoS routing protocol in which a location-delay prediction scheme, based on a location-resource update protocol has been performed. The location updates contain resource information pertaining to the node sending the update. This resource information for all nodes in the network and the location prediction mechanism are together used in the QoS routing decisions. There are dynamic changes in topology and resource availability due to the high degree of mobility of nodes in the ad hoc network. Due to these changes, the topological and routing information used by current network protocols is rendered obsolete very quickly. The advantage of this system is the prediction of new location based on previous location is made when there is variation in the geographical location. QoS routing based on the resource availability at the intermediate nodes in the source to destination route is performed which is rare in other location based routing scheme. But, accurate prediction on velocity and direction is not made when there are dynamic changes in the direction. The transmission is made only in linear pattern (i.e., angular velocity is kept as zero).

### 6.5 QoS Multicast Routing Protocol with Dynamic group topology (QMRPD)
The QMRPD is a hybrid protocol which attempts to significantly reduce the overhead of constructing a multicast tree with multiple QoS constraints. In QMRPD, a multicast group member can join or leave a multicast session dynamically, which should not disrupt the multicast tree. It satisfies the multiple QoS constraints and least cost's (or lower cost) requirements. Its main objective is to construct a multicast tree that optimizes a certain objective function (e.g., making effective use of network resources) with respect to performance-related constraints (e.g., end-to end delay bound, inter-receiver delay-jitter bound, minimum bandwidth available, and maximum packet-loss probability) and design a multicast routing protocol with dynamic group topology. It attempts to minimize the overall cost of the tree. The dynamic group membership has been handled by this protocol with less message processing overhead.

### 6.6 QoS Optimized Link State Routing (QOLSR)
The Optimized Link State Routing (OLSR) protocol [23] is a proactive link state routing protocol for MANETs. One key idea is to reduce control overhead by reducing the number of broadcasts as compared with pure flooding mechanisms. The basic concept to support this idea in OLSR is the use of multipoint relays (MPRs) [23, 27]. MPRs refer to selected routers that can forward broadcast messages during the flooding process. To reduce the size of broadcast messages, every router declares only a small subset of all of its neighbors. "The protocol is particularly suitable for large and dense networks" [23]. MPRs act as intermediate routers in route discovery procedures. Hence, the path discovered by OLSR may not be the shortest path. This is a potential disadvantage of OLSR.

OLSR has three functions: packet forwarding, neighbor sensing, and topology discovery. Packet forwarding and neighbor sensing mechanisms provide routers with information about neighbors and offer an optimized way to flood messages in the OLSR network using MPRs. The neighbor sensing operation allows routers to diffuse local information to the whole network. Topology discovery is used to determine the topology of the entire network and calculate routing tables. OLSR uses four message types: Hello message, Topology Control (TC) message, Multiple Interface Declaration (MID) message, and Host and Network Association (HNA) message. Hello messages are used for neighbor sensing. Topology declarations are based on TC messages. MID messages contain multiple interface addresses and perform the task of multiple interface declarations. Since hosts that have multiple interfaces connected with different subnets, HNA messages are used to declare host and associated network information. Extensions of message types may include power saving mode, multicast mode, etc.

### 6.7 Ad hoc QoS on-demand routing (AQOR)
This protocol uses limited flooding to discover the best route available in terms of smallest end-to-end delay with bandwidth guarantee. A route request packet includes both bandwidth and end-to-end delay constraints. Let Tmax denote the delay constraint. If a node can satisfy both constraints, it will rebroadcast the request to the next hop and switch to explore status for a short period of 2Tmax. If multiple request packets arrive at the destination, it will send back a reply packet along each of these routes. Intermediate nodes will only forward the reply, if they are still in explored state. However, the bandwidth reservation for each flow is only activated by the arrival of the first data packet from the source node. Delay is measured during route discovery. The route with the least delay is chosen by the source. No mechanism for connection tear-down is needed or integrated, since all reservations are only temporary. Timers are reset every time a route is used. So there is an upper time bound after which broken routes are detected. To further reduce communication overhead during route discovery, AQOR can work with some location aided routing protocols. For delay violation detection, the estimated time offset between the systems clocks of source and destination node has to be known.

## 7. SUMMARY OF QOS ROUTING PROTOCOLS

To facilitate a comparison among the different QoS-aware routing protocols, the salient features of the QoS routing protocols is described in a table. The table lists the design constraints listed earlier such as Route discovery, Resource reservation, Route maintenance, QoS metrics constrained, Network architecture and routing overhead and discussing how each protocol addresses.

| Routing protocol | Network architecture | Route discovery | Type of QoS guarantee | Resource reservation | QoS metrics | Routing overhead |
|---|---|---|---|---|---|---|
| CEDAR | Hierarchical | Proactive/ Reactive | Soft | Yes | Bandwidth | core setup |
| MRP | Hierarchical | Reactive | Soft | Yes | Bandwidth | Full flooding of RREQ |
| GAMAN | Hierarchical | Reactive | Soft | Yes | Bounded delay, packet loss rate | Node traversal delay |
| PLBQR | Location prediction | Proactive/ Reactive | Soft | No | Delay, and Bandwidth | Route recomputation in anticipation of link breakage |
| QMRPD | Hierarchical | Reactive | Pseudo-hard | Yes | Bandwidth, Delay, Delay-jitter and cost | Less message processing overhead |
| QOLSR | Hierarchical | Proactive | Soft | Yes | Throughput and Delay | Minimum flooding of RREQ |
| AQOR | Flat | Reactive | Soft | Yes | Bandwidth, Delay | Full flooding of RREQ |
| TBR | Flat | Reactive | Soft | Yes | Bandwidth, Delay | Minimum flooding of RREQ |
| QAODV | Flat | Reactive | Soft | No | Bandwidth, Delay | Node traversal delay |

## 8. REFERENCES

[1] C.R.Lin and J.S. Liu., "QoS routing in ad hoc wireless networks", IEEE J.Select.Areas Commun.,vol.17, pp.1488-1505, 1999.
[2] C.Perkins, "Ad-hoc On-Demand Distance Vector (AODV) routing", RFC3561[S], 2003.
[3] D.B.Johnson, D.A.Maltz, Y.C.Hu, "The Dynamic Source Routing protocol for mobile ad hoc networks",Internet Draft, 2004.
[4] J.Hong, "Efficient on-demand routing for mobile ad hoc wireless access networks", IEEE journal on selected Areas in Communications 22(2004), 11-35.
[5] Luo Junhai, Xue Lie and Ye Danxia "Research on multicast routing protocols for mobile ad hoc networks" Computer Networks52(2008), 988-997.
[6] S. Chen, Routing Support for Providing Guaranteed End-to-End Quality-of-Service, Ph.D Thesis,University of IL at Urbana-Champaign, 1999.
[7] S. Chakrabarti and A. Mishra, "QoS issues in ad-hoc wireless networks" IEEE Commun. Mag., vol.39,pp.142-148, Feb. 2001.
[8] J.N. Al-Karaki and A.E.Kamal, "Quality of Service routing in mobile ad hoc networks: Current and future trends" in Mobile Computing Handbook, CRC Publishers, 2004.
[9] T.B.Reddy I.Karthigeyan, B.Manoj and C.S.R.Murthy, "Quality of service provisioning in ad hoc wireless networks: a survey of issures and solutions." Vol.4, pp.83-124, 2006.
[10] S.Saunders, "Antennas and Propagation for Wireless Communication System Concept and design",New York, USA: John Wiley and Sons, 1999.
[11] Y.Luo, J.Wang and J.Chen et al., "Algorithm based on mobility prediction and probability for energy efficient multicasting in ad-hoc networks" Computer research and development vol.43(2), pp.231-237,2006.
[12] Y.Yang and R.Kravets "Contention-aware admission control for ad hoc networks", IEEE Trans.Mobile Comput., vol.4, 363-377, 2005.
[13] C.R.Lin and J.S. Liu., "QoS routing in ad hoc wireless networks", IEEE J.Select.Areas Commun.,vol.17, pp.1426-11438, 1999.
[14] S.Chen and K.Nahrstedt, "Distributed quality-of-service routin in ad hoc networks" IEEE J.Select.Areas Commun., vol.17, pp.1488-1505, 1999.
[15] A.R.Bashandy, E.K.P.Chong and A.Ghafoor, "Generalized quality-of-service routing with resource allocation" IEEE J.Select.Areas Commun., vol.23, pp.450-463, 2005.
[16] A.Abradou and W.Zhuang, "A position based QoS routing scheme for UWB mobile ad hoc networks" IEEE J.Select.Areas Commun., vol.24, pp.850-856, 2006.
[17] L.Li, C.Li, "A hierarchical QoS multicast routing protocol for mobile ad-hoc networks", Chinese Journal of Electronics 15(4) (2006), 573- 577.
[18] M.S.Kumar, C.Venkatesh, A.M.Ntarajan, "Performance comparison of multicast protocol for physically hierarchical ad-hoc networks using neural Concepts", 7th International conference on Signal Processing Proceedings ICSP, 2004, 1581-1584.
[19] R.Sivakumar, P.Sinha and V.Bharghavan, "CEDAR: a core-extraction distributed ad hoc routing algorithm" IEEE J.Select.Areas Commun., vol.17, pp.1454-1465, 1999.
[20] L.A.Latif, A.Alliand, C.C. Ooi, "Location based geocasting and forwarding (LGF) routing protocol in mobile-adhoc network", Proceedings on the Advanced Industrial Conference on Telecommunications/Service assurance with partial Intermittent Resources (2005) 536-541.
[21] K.Chen, K.Nahrstedt, "Effective location guided-tree construction algorithms for small group multicast in MANET", Proceedings of the INFOCOM (2002) 1180-1189.
[22] M.Transier, H.Fubler, J.Widmet et al., "Scalable position based multicst for mobile ad-hoc networks" in:Proceddings of the First International; workshop on Broadband wireless multimedia: Algorithms, Arcgutectures, and applications, San Jose, CA, Oct. 2004.
[23] S. Basagni, I. Chlamtac, V. R. Syrotiuk, B. A. Woodward A, "A Distance Routing Effect Algorithm for Mobility (DREAM)", In Proc. ACM/IEEE Mobicom, pages 76-84, October 1998.
[24] M. Amr M. Abd El Moniem "An Intelligent cryptographic Algorithmic system To Secure Computer Networks" PhD Thesis, Cairo University, 2001.
[25] Network Working Group, "Routing Protocol Performance Issues and Evaluation Considerations", RFC 2501.
[26] Rahul Jain, Anuj Puri and Raja, "Geographical routing using partial information for wireless ad hoc networks", in IEEE Personal Communications, pages 48-57, February 2001
[27] Prince Samar, Stephen B. Wicker, "On the behavior of communication links of a node in a multi-hop mobile environment", MobiHoc 2004: 145-156
[28] Sunil Taneja and Ashwani Kush, "A Survey of Routing Protocols in Mobile Ad Hoc Networks" International Journal of Innovation, Management and Technology, Vol. 1, No. 3, August 2010 ISSN: 2010-0248.

**AUTHORS PROFILE**

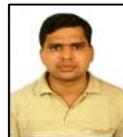
**Sanjeev Gangwar** is Assistant Professor in the Department of Computer Applications, VBS Purvanchal University, Jaunpur, India. He obtained his M.C.A degree from Rohilkhand University Bareilly and M.Phil. in Computer Science from Vinayaka mission University, Tamilnadu. He is currently doing research in Mobile Ad-Hoc Network.

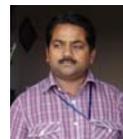
**Saurabh Pal** received his M.Sc. (Computer Science) from Allahabad University, UP, India (1996) and obtained his Ph.D. degree from the Dr. R. M. L. Awadh University, Faizabad (2002). He then joined the Dept. of Computer Applications, VBS Purvanchal University, Jaunpur as Lecturer. At present, he is working as Head and Sr. Lecturer at Department of Computer Applications.

Saurabh Pal has authored a commendable number of research papers in international/national Conference/journals and also guides research scholars in Computer Science/Applications. He is an active member of IANEG, IACSIT CSI, Society of Statistics and Computer Applications and working as Editor/Editorial Board Member/Reviewer for more than 20 international journals. His research interests include Image Processing, Data Mining, Grid Computing and Artificial Intelligence.

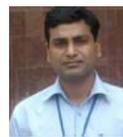
**Krishan Kumar** is Assistant Professor in the Department of Computer Applications, Gurukula kangri Vishwavidyalaya ,Haridwar, India. He obtained his M.C.A degree from IMS GHAZIABAD and Ph. D. in Computer Science and Information Technology from M.J.P.Rohilkhand University, Bareilly.

Krishan Kumar has authored 04 research papers in international/national Conference/journals and also guides research scholars. He is an active member of various international societies and working as reviewer for 2 international journals.